\documentclass[a4paper]{article}

\usepackage{INTERSPEECH2022}
\usepackage{cite} 
\usepackage{hyperref}

\title{Style-Label-Free: Cross-Speaker Style Transfer by Quantized VAE and Speaker-wise Normalization in Speech Synthesis}
\name{Chunyu Qiang$^*$, Peng Yang$^*$, Hao Che, Xiaorui Wang, Zhongyuan Wang}
\address{
 Kwai, Beijing, P.R. China}
\email{\{qiangchunyu, yangpeng, chehao, wangxiaorui, wangzhongyuan\}@kuaishou.com}

\begin{document}
\maketitle
\begin{abstract}
  Cross-speaker style transfer in speech synthesis aims at transferring a style from source speaker to synthesised speech of a target speaker's timbre. Most previous approaches rely on data with style labels, but manually-annotated labels are expensive and not always reliable. In response to this problem, we propose Style-Label-Free, a cross-speaker style transfer method, which can realize the style transfer from source speaker to target speaker without style labels. Firstly, a reference encoder structure based on quantized variational autoencoder (Q-VAE) and style bottleneck is designed to extract discrete style representations. Secondly, a speaker-wise batch normalization layer is proposed to reduce the source speaker leakage. In order to improve the style extraction ability of the reference encoder, a style invariant and contrastive data augmentation method is proposed. Experimental results show that the method outperforms the baseline. We provide a website with audio samples\href{https://qiangchunyu.github.io/UGM_DEMO/UGMDEMO.html}{$^1$}.

\end{abstract}
\renewcommand{\thefootnote}{\fnsymbol{footnote}} 
\footnotetext[1]{These authors contributed equally to this work.} 
\footnotetext[2]{https://qiangchunyu.github.io/UGM\_DEMO/UGMDEMO.html} 


\rfoot{Page \thepage}

\section{Introduction}

With the development of deep learning, speech synthesis technology has rapidly advanced\cite{sotelo2017char2wav, wang2017tacotron,peng2020non,kim2020glow,liu2021vara,elias2021parallel}. Improving the expressiveness and controllability of TTS systems for a better listening experience has attracted more attention and research.Most of the traditional cross-speaker style transfer methods require style labels to assist in transforming the speaking style of the source speaker into the synthesized speech of the target speaker's timbre.



Many cross-speaker style transfer models have been proposed, most of which require all \cite{whitehill2019multi, li2021controllable, kang2021unitts, pan2021cross} or part \cite{hsu2018hierarchical, habib2019semi, wu2021cross} of style labels, which are expensive to construct, and not always reliable. Some expressive and controllable speech synthesis methods that does not require style labels have been proposed \cite{skerry2018towards, wang2018style, zhang2019learning, lee2021styler}, but they lack interpretability and can only achieve intra-speaker style control, which is difficult to achieve cross-speaker style transfer. The widely used reference encoder methods are based on global style tokens(GST)\cite{wang2018style} or variational autoencoders (VAEs)\cite{kingma2013auto,zhang2019learning, kenter2020improving, kenter2019chive, hsu2018hierarchical, habib2019semi}. VAE is used to model the variance information in the latent space with Gaussian prior as a regularization, which does not require explicit annotations. Sun et al. propose a sequential prior in a discrete latent space using vector quantization (VQ) \cite{sun2020generating}. Habib et al. propose semi-supervised learning method to learn the latent of VAE model \cite{habib2019semi}.Hsu et al. propose Gaussian mixture VAE models to disentangle different attributes \cite{hsu2018hierarchical}. The above methods are difficult to disentangle speaker timbre and style information without style labels. Many methods use intercross training\cite{bian2019multi}, gradient reversal, domain adversarial training \cite{ sorin2020principal, ganin2016domain, zhang2021denoispeech,lee2021styler, zhou2021seen} or add multiple loss functions\cite{xue2021cycle, an2021improving, kulkarni2021improving, joo2020effective, li2021controllable, whitehill2019multi, nachmani2018fitting} to better reduce the source speaker leakage. Li et al. propose a controllable emotional transfer method by adding an emotion classifier with feedback cycle\cite{li2021controllable}. Whitehill et al. propose a contrastive cycle consistency training scheme with paired and unpaired triplets to ensure the use of information from all style dimensions\cite{whitehill2019multi}. These methods require style labels for style classification or to construct paired/unpaired triplets. Therefore, cross-speaker style transfer faces great challenges without style labels. 




This paper focuses on cross-speaker style transfer in speech synthesis without style labels. Due to speaker timbre information and style information in speech are highly entangled, the key to solving this problem is how to build a high-performance style extractor and effectively reduce the source speaker leakage. Moreover, it is necessary to clearly separate the latent space of different style information without style labels. Instead of existing methods that require style labels, a cross-speaker style transfer method Style-Label-Free is proposed, which can realize the transfer of styles from source speaker to target speaker without style labels. This paper demonstrates the effectiveness of the proposed methods on the global style embedding, and the future work will focus on extending to fine-grained style embedding.

\begin{figure*}[t]
 \centering
 \includegraphics[width=\linewidth]{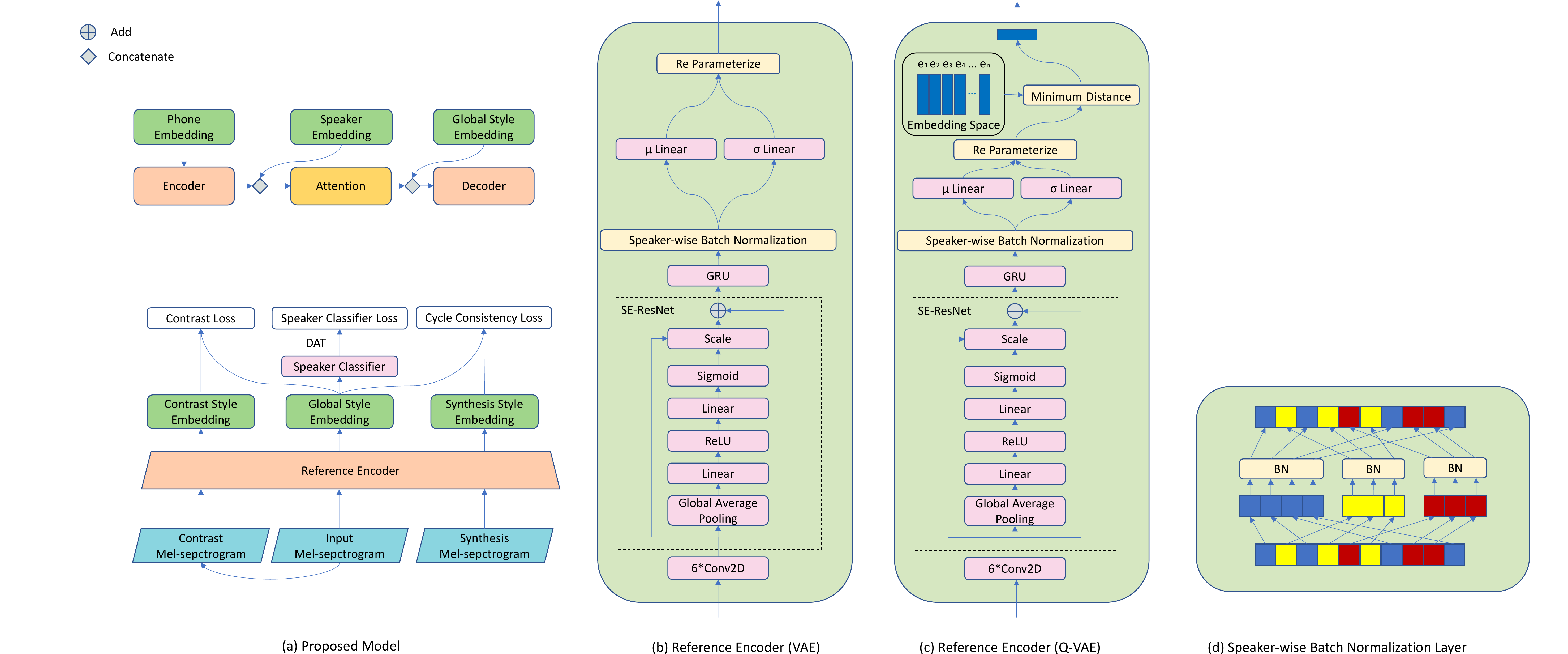}
 \caption{The architecture of (a) proposed model, (b) Reference Encoder (VAE), (c) Reference Encoder (Q-VAE), (d) Speaker-wise Batch Normalization Layer.}
 \label{fig:speech_production}
\end{figure*}

The contributions of this paper are as follows.

\begin{itemize}
\item A reference encoder structure based on quantized variational autoencoder (Q-VAE) and style bottleneck is designed to extract discrete style representations.
\end{itemize}

\begin{itemize}
\item A speaker-wise batch normalization layer is proposed to reduce the source speaker leakage in cross-speaker style transfer. 
\end{itemize}

\begin{itemize}
\item A style invariant and contrastive data augmentation method is proposed to improve the style extraction ability of the reference encoder. 
\end{itemize}


\section{Method}

The proposed framework is illustrated in Figure 1(a). As shown, the proposed model is an attention-based seq2seq framework, Tacotron-like systems take a text sequence, a speaker id and a reference acoustic features as input, and use autogressive decoder to predict a sequence of acoustic features frame by frame. Meanwhile, style paired/unpaired triples are constructed by style invariant and contrastive data augmentation method to compute contrastive cycle consistency loss. As shown in Figure 1(b)(c), two reference encoder structures based on VAE and Q-VAE are designed to extract style information. In order to alleviate the highly entangled problem in cross-speaker style transfer and improve the style extraction ability of the model, a style bottleneck sub-network\cite{pan2021cross} is introduced to the reference encoder. The style bottleneck network consists of 6 layers 2D convolutional networks and a (Squeeze-and-Excitation based ResNet architecture) SE-ResNet block \cite{hu2018squeeze}. The SE-ResNet block can adaptively recalibrate channel-wise feature responses by explicitly modelling interdependencies among channels, and produce significant performance improvements. The reference encoders use the speaker-wise batch normalization layer described in Figure 1(d) to reduce the source speaker leakage. 

\subsection{Variational Autoencoder} \label{variance}
\subsubsection{VAE} \label{VAE}
The model obtains a continuous and complete latent space distribution of styles through the VAE \cite{kingma2013auto} structure to improve the style control ability. As illustrated in Figure1(b), the variational layer takes the last output of the GRU layer through the speaker-wise batch normalization layer as input, and then inputs the two fully connected layers (mean Linear and variance Linear) to obtain the mean and variance of the multivariate Gaussian distribution. Finally, a 64-dimensional vector is sampled from this Gaussian distribution as input to the decoder (concatenated with the Pre-Net output at each step).

In the rest of this paper, $D_{KL}$ is referred to as KL loss, $\mathcal{N}(\cdot)$ is referred as Gaussian distribution and $({\hat{\mu}}, {\hat{\sigma}})$ is referred as the $(mean, variance)$ of style latent space distribution. Random operations in the network cannot be processed by backpropagation, "reparameterization trick" is introduced to VAE: $\boldsymbol{z} = {\hat{\mu}} + {\hat{\sigma}} \odot \phi ; \phi \sim \mathcal{N}(0, I)$.
During the training process, KL loss is easily reduced to zero, which is called KL collapse. Three tricks are used to solve this problem. Firstly, the KL annealing is introduced. Secondly, a staged optimization method is adopted to optimize the reconstruction loss first and then the KL loss. Finally, a margin $\Delta$ is introduced to limit the minimum value of the kl loss as shown in Equation (1) shown.

\begin{equation}
 \mathcal{L}_{kl} = max(0, D_{KL}[\mathcal{N}({\hat{\mu}},{\hat{\sigma}}^2)||\mathcal{N}(0, I)]-\Delta)
 \label{KL}
\end{equation}

\begin{figure*}[t]
 \centering
 \includegraphics[width=\textwidth,height=0.27\textwidth]{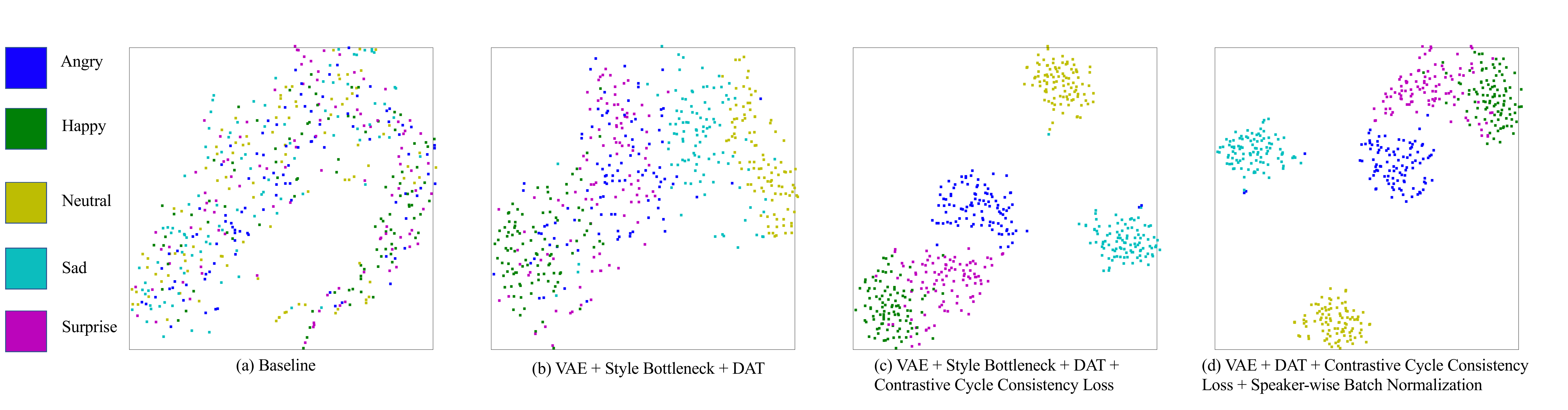}
 \caption{t-SNE plot of global style embeddings for 500 style balanced utterances}
 \label{fig:t-sne}
\end{figure*}

\subsubsection{Q-VAE}

Inspired by existing work\cite{sun2020generating}, discretizing the latent features using vector quantization (VQ) can generate more naturally sounding samples, and we propose quantized variational autoencoders (Q-VAE). The Q-VAE quantifies the VAE output into a fixed number of classes. Meanwhile, the quantized representation from the continuous latent space ensures reasonable diversity across samples. To distinguish it from VQ-VAE we call this structure Q-VAE.

As shown in Figure 1(c), Q-VAE extends the VAE by adding a discrete codebook component to the network. The output of the VAE is compared with all the vectors in the codebook, and the codebook vector closest in euclidean distance is fed into the decoder. The vector quanatization loss consists of two parts: the commitment loss (get the VAE output to commit as much as possible to its closest codebook vector)of Equation (2) and the codebook loss (get the chosen codebook vector as close to the VAE output as possible) of Equation (3). $\mathcal{E}(\cdot) $ is referred as the reference encoder, $sg[\cdot] $ stands for "stop gradient", $\boldsymbol z$ is referred as VAE output, and $\boldsymbol e$ is referred as the codebook vector.

\begin{equation}
 \mathcal{L}_{commitment} = ||\boldsymbol z - sg[\boldsymbol e]||^2_2
 \label{L_commitment}
\end{equation}

\begin{equation}
 \mathcal{L}_{codebook} = ||sg[\boldsymbol z] - \boldsymbol e||^2_2
 \label{L_codebook}
\end{equation}

In order to get faster convergence speed, exponential moving averages(EMA)\cite{van2017neural} is used instead of codebook loss.




\subsection{Speaker-wise Batch Normalization Layer}
The speaker timbre and style in speech signals are highly entangled, and reducing the source speaker leakage plays an important role in the task of cross-speaker style transfer. Therefore, a speaker-wise batch normlization layer is proposed to solve this problem. As illustrated in Figure1(d), the method is to normalize the vectors belonging to the same speaker in each batch, and each speaker stores a set of batch normalization parameters separately. In Equation (4) (5), $({\mu_S}, {\sigma_S^2})$ represents the $(mean, variance)$ of a single speaker, $\boldsymbol {V_s}$ is the set of vectors belonging to the same speaker in the current batch, and $m$ represents the number of vectors it contains.

\begin{equation}
  {\mu_S} = \frac {1}{m}\sum\limits_{v\in{\boldsymbol {V_s}}}{v}
 \label{mu}
\end{equation}

\begin{equation}
 {\sigma_S}^2 =  \frac {1}{m}\sum\limits_{v\in{\boldsymbol{V_s}}}({v} - \mu_S)^2
 \label{sigma}
\end{equation}

The input features are k-dimensional and $\theta$ is a small positive constant to prevent numerical instability. The proposed speaker-wise batch normalization for each dimension is defined as:

\begin{equation}
 \widetilde{{v}}^{(k)} = \frac {{v}^{(k)} - {\mu_S}^{(k)}} {\sqrt[]{{{\sigma_S}^{(k)}}^2 + \theta}}
 \label{SWBN}
\end{equation}

Simply normalizing each input to a layer may change what the layer represents. Additional learnable parameters $\omega$ and $\lambda$ are introduced for scaling and moving the normalized activations to enhance the representational power of the layer. 

\begin{equation}
 \mathcal{SN}({v}^{(k)}) = \omega^{(k)}\widetilde{{v}}^{(k)} + \lambda^{(k)}
 \label{SN}
\end{equation}


\subsection{Style Invariant and Contrastive Data Augmentation}
Existing methods have demonstrated that contrastive cycle consistency loss is effective for style transfer\cite{whitehill2019multi, nachmani2018fitting}. But in these methods, paired/unpaired triplets cannot be constructed without style labels.

\subsubsection{Style Invariant}
The ground truth acoustic features $\boldsymbol y$ and the synthesized acoustic features $\boldsymbol{\hat{y}}$ constitute paired two-tuples to compute cycle consistency loss. Due to teacher-forcing, these two features are almost the same, which leads to overfitting of the reference encoder. In order to solve this problem, a style invariant data augmentation method is proposed to enhance model robustness. The acoustic features are randomly clipped with a window of length 300 frames at each step as the input to the reference encoder. 

\begin{equation}
 \mathcal{L}_{cycle} = \frac{1}{n^2} \sum(\mathcal{E}(\boldsymbol y)^T*\mathcal{E}(\boldsymbol y) - \mathcal{E}(\boldsymbol{\hat{y}})^T*\mathcal{E}(\boldsymbol{\hat{y}}))^2
 \label{L_cycle}
\end{equation}

\subsubsection{Style Contrastive }

 Studies have found that the style information of a single speaker is highly correlated with pitch, energy and duration\cite{xie2021multi}. Therefore, a method is proposed to augment style contrastive data $\boldsymbol{\check{y}}$ by randomly modifying the pitch, energy and duration of speech within a certain range. We expect the computed embeddings of $\boldsymbol y$ and $\boldsymbol{\check{y}}$ to be different, and $\boldsymbol{\check{y}}$ is obtained by augmenting $\boldsymbol y$, the computed embedding distance should not be too far, so a margin $\Gamma$ is introduced to limit the minimum value of the  contrastive loss. 
\begin{equation}
\begin{aligned}
 \mathcal{L}_{contrast} = max(0, \Gamma - \frac{1}{n^2} \sum(\mathcal{E}(\boldsymbol y)^T*\mathcal{E}(\boldsymbol y) - \\ \mathcal{E}(\boldsymbol{\check{y}})^T*\mathcal{E}(\boldsymbol{\check{y}}))^2 )
 \label{L_constrast}
\end{aligned}
\end{equation}

 As described in Equation (8) (9), the cycle consistency loss and  contrastive loss are calculated using the Gram matrix that can capture the local statistics of the audio signal in the frequency and time domain.\cite{ma2018neural}

\subsection{Training Details}
The model uses a gradient reversal layer(GRL) for adversarial speaker training. As shown in Figure 1(a), the extracted global style embedding is fed into the speaker classifier, which consists of a fully connected layer, a softmax layer and a GRL. The speaker classification loss is denoted by $\mathcal{L}_{spk}$.

The total loss of the model without noise modeling is:
\begin{equation}
\begin{aligned}
 \mathcal{L} = \mathcal{L}_{spec}+\mathcal{L}_{stop}+\alpha\mathcal{L}_{kl}+\beta\mathcal{L}_{spk}+\\ \gamma\mathcal{L}_{cycle}+\delta\mathcal{L}_{contrast} + \zeta\mathcal{L}_{commitment}
 \label{L}
\end{aligned}
\end{equation}

where [$\alpha, \beta, \gamma, \delta, \zeta$] are the weights of [$\mathcal{L}_{kl}, \mathcal{L}_{spk}, \mathcal{L}_{cycle}, \mathcal{L}_{contrast}, \mathcal{L}_{commitment}$]. [$\mathcal{L}_{spec}, \mathcal{L}_{stop}$] represent reconstruction loss and stop token loss. In order to make the model converge effectively, a staged training method is adopted, and the optimization order is $\mathcal{L}_{spec}, \mathcal{L}_{stop}, \mathcal{L}_{commitment}, \mathcal{L}_{kl}, \mathcal{L}_{spk}, \mathcal{L}_{cycle}, \mathcal{L}_{contrast}$.

\begin{table*}[th]
 \caption{Comparison of our proposed method}
 \label{tab:example}
 \centering
 \begin{tabular}{l|ccc|ccc}
  \toprule
\multicolumn{1}{c|}{{w/o}} & \multicolumn{3}{c|}{Style Similarity MOS} & \multicolumn{3}{c}{Speaker Similarity MOS} \\ \cline{2-7} 

\multicolumn{1}{c|}{}           & Baseline   & VAE   & Q-VAE     & Baseline   & VAE   & Q-VAE      \\ \hline
Speaker DAT                 & \textbackslash{} & 3.01 $\pm$0.051 & 3.30 $\pm$0.071     & \textbackslash{} & 3.63 $\pm$0.043 & 3.74 $\pm$0.073     \\
Contrastive Cycle Consistency Loss                & \textbackslash{} & 2.90 $\pm$0.094 & 3.19 $\pm$0.084     & \textbackslash{} & 3.86 $\pm$0.080 & 3.87 $\pm$0.015     \\
Speaker-wise Batch Normalization                   & \textbackslash{} & 2.87 $\pm$0.060 & 2.99 $\pm$0.050     & \textbackslash{} & 3.54 $\pm$0.065 & 3.70 $\pm$0.041     \\ \hline
All                    & 2.51 $\pm$0.032    & 2.53 $\pm$0.091 & 2.94 $\pm$0.081     & 3.10 $\pm$0.076    & 3.43 $\pm$0.078 & 3.65 $\pm$0.065     \\ \hline
None                   & \textbackslash{} & 3.19 $\pm$0.026 & \textbf{3.42 $\pm$0.045} & \textbackslash{} & 3.86 $\pm$0.098 & \textbf{3.87 $\pm$0.092} \\ \hline
  
 \end{tabular}
 
\end{table*}

\section{Experiments}
\subsection{Experimental Step}
\subsubsection{Database}
An open source multi-speaker emotional speech dataset (ESD)\cite{zhou2021seen} is used with only 10 native Mandarin speakers (5 males and 5 females), two of which contained only natural style data (as the target speaker timbre for the experiment) and the others contained all emotions. For the cross-speaker style transfer task, no style labels are used during training and inference.The dataset contains 350 parallel utterances with an average duration of 2.9 seconds per speaker, all speech waveforms sampled at 16kHz are converted to mel-spectrogram with a frame size of 960 and hop size of 240.

\subsubsection{Compared Models}

\begin{itemize}
\item Baseline: \cite{zhang2019learning} is used as our baseline, and add a speaker embedding layer structure to this model. 
\end{itemize}
\begin{itemize}
\item VAE: Proposed model described in Sec 2.1.1.
\end{itemize}

\begin{itemize}
\item Q-VAE: Proposed model described in Sec 2.1.2.
\end{itemize}

\begin{figure}[t]
 \centering
\includegraphics[width=0.5\textwidth,height=0.35\textwidth]{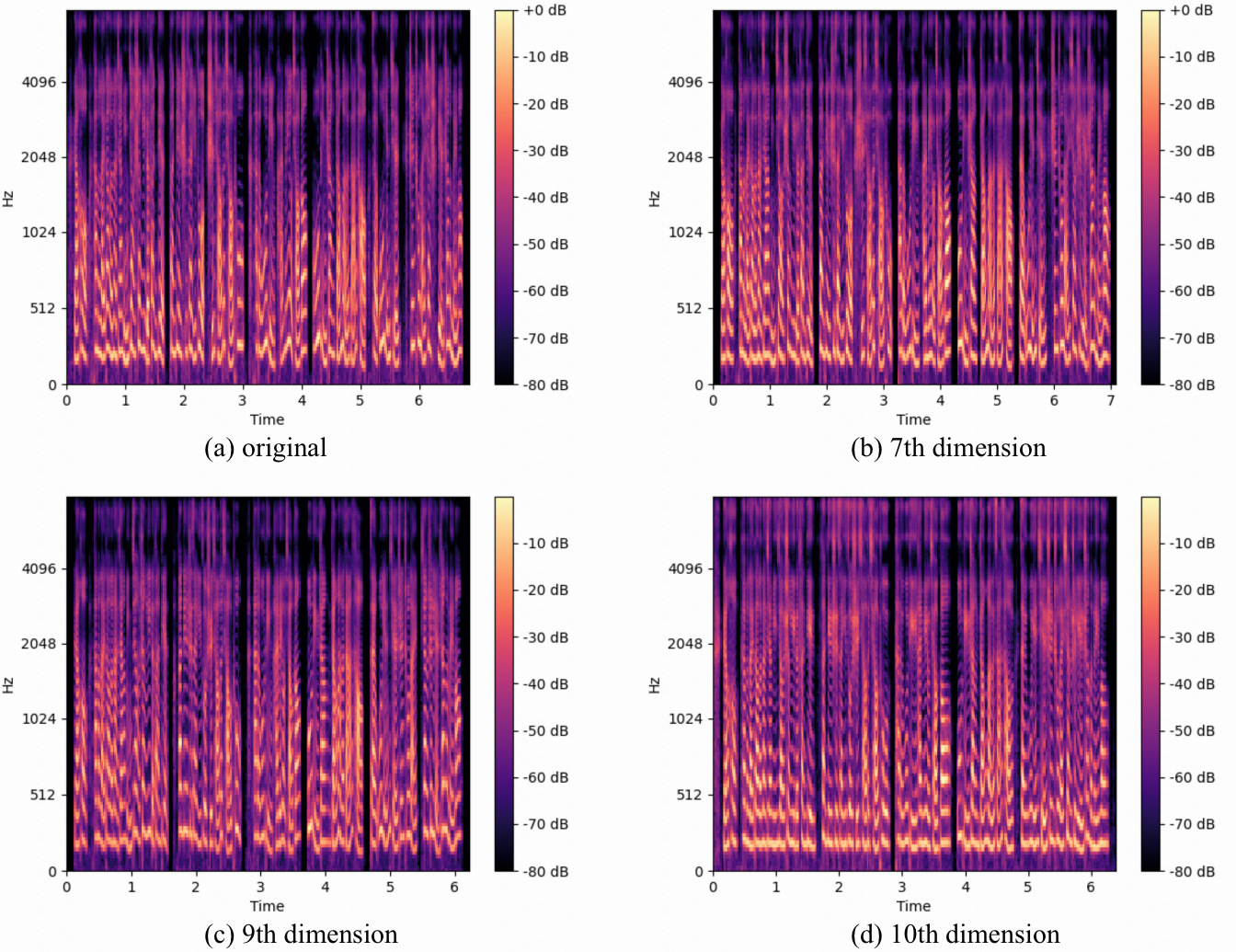}
 \caption{Mel-spectrogram of single-dimensional control of global style embedding: (a) is original sample, (b) changes the duration by controlling dimension 7, (c) changes the energy by controlling dimension 9, (d) changes the pitch by controlling dimension 10.}
 \label{fig                                   :speech_production}
\end{figure}

\subsection{Results}

\subsubsection{Objective Results}
To objectively compare the style clustering ability of several proposed methods, 500 utterances are randomly selected from different speakers, where the number of each style is balanced. Since the global style embedding extracted by Q-VAE reference encoder is a fixed number of discrete points, VAE reference encoder is used for comparison. The 64-dimensional global style embeddings are reduced to 2-dimensional vector using t-SNE\cite{van2008visualizing} and plotted in Figure 2. In the multi-speaker multi-style task, as shown in Figure 2(a), the VAE reference encoder structure cannot extract the speaker-independent style information. Speaker information and style information are highly entangled without style label, which is consistent with the existing conclusions. As shown in Figure 2(b), the methods of style bottleneck and speaker DAT can make the style vector have a weak clustering effect, but the boundary of the latent space is not clear enough. The proposed style invariant and contrastive data augmentation method to construct contrastive cycle consistency loss can improve the style extraction ability of the reference encoder, as shown in Figure 2(c)(d), the model can already perform clustering with clear boundaries in the latent space. Figure 3 shows the mel-spectrogram synthesized by controlling different single dimensions of the global style embedding, demonstrate the effect of our proposed methods. 

\subsubsection{Subjective Results}

In the cross-speaker style transfer task, a good model should preserve the timbre of the speaker labels while preserving the style similar to the reference audio. Therefore, the model is evaluated using two metrics, style similarity and speaker similarity, which refer to the similarity in expected speaking style and timbre between natural speech and synthesized speech. The two similarities are evaluated using a mean opinion score (MOS) evaluation using a human scoring experiment. An ablation study is performed by comparing the proposed method with several variants achieved by removing one or all structures. The result are shown in Table 1.

In terms of speaker similarity MOS, both VAE and Q-VAE methods have achieved acceptable results. Compared with baseline, speaker DAT and speaker-wise batch normalization structures can better reduce the source speaker leakage. Since Q-VAE outputs a fixed number of style cluster centroids, it has better discreteness, so it achieves the best speaker similarity. 

In terms of style similarity MOS, baseline cannot achieve cross-speaker style transfer. Since the teacher-forcing and the current frame of the autoregressive structure is dependent on the previous frame during the training process, the model tends to ignore the style embedding. Q-VAE gives the model more explicit and discrete style information, and methods such as speaker-wise batch normalization and contrastive cycle consistency loss can reduce the source speaker leakage, so it achieves a better style similarity.

\section{Conclusions}

In this paper, Style-Label-Free, a cross-speaker style transfer method is proposed. 
The proposed methods such as Q-VAE, style invariant and contrastive data augmentation, and speaker-wise batch normalization, build a high-performance style extractor and effectively reduce the source speaker leakage. Experiments show that the effectiveness of our proposed methods. The future work will focus on extending the proposed methods to fine-grained style control.





\end{document}